\documentclass[doublespacing ]{elsart}
\usepackage{graphicx}
\usepackage{amssymb}

\newcommand{\svar}{T \ln(t/\tau_0)}

\newcommand{\beq}{\begin{equation}}
\newcommand{\eeq}{\end{equation}}
\newcommand{\bfig}{\begin{figure}}
\newcommand{\efig}{\end{figure}}

\begin{document}
%\draft
\begin{frontmatter}
\title{Effects of the Magnetic Field on the Relaxation of Small Particle Systems}
\author{\`Oscar Iglesias and Am\'{\i}lcar Labarta}
\address{Department de F\'{\i}sica Fonamental, Facultat de F\'{\i}sica, Universitat\\
de Barcelona, Diagonal 647, 08028 Barcelona, Spain}
\begin{abstract}
We study the effect of a magnetic field on the thermal relaxation of non-interacting
small monodomain particle systems particles with a distribution of anisotropy constants and random easy-axes directions. 
Numerical calculations of the relaxation curves for different distribution widths, 
and under different magnetic fields $H$ and temperatures $T$, have been performed in the 
framework of a two-state approximation. We show how the obtained data can be analyzed in terms 
of an modified $Tln(t/\tau_0)$ scaling from which the field dependence of the mean relaxing 
energy barriers can be extracted, a microscopic information which is not easily obtainable by other methods.
\end{abstract}

\end{frontmatter}
%_/_/_/_/_/_/_/_/_/_/_/_/_/_/_/_/_/_/_/_/_/_/_/_/_/_/_/_/_/_/_/_/_/_/_/_/_/_/_/ 
\section{Introduction}

\label{introduction} 

The progressive miniaturization of the elementary recording units for magnetic 
recording has put forward the problem of stability of magnetization against thermal 
activation. \cite{Dormannadv97,JMMM221} Therefore, the study of the thermal relaxation of the magnetization 
in fine particle systems is of crucial interest to ascertain the stability of the 
recorded information. Although the first studies in this field date back to
the 40's \cite{Stoner48,Neel2,Street}, at present, there is no complete model
to account for all the experimental findings. 
One of the points that has not been completely clarified is the influence of 
a magnetic field in the relaxation.

We want to account for the experimental
studies on the relaxation of small-particle systems, which essentially
measure the acquisition of magnetization of an initially demagnetized sample
under the application of a magnetic field. \cite
{Vincentjpe94,Kimjm99,Lisfijm99,Mayergoyzjap99,Montseprb99} In this kind of
experiments, the field modifies the energy barriers of the system that are
responsible for the time variation of the magnetization, as well as the
final state of equilibrium towards which the system relaxes.
Here, we will demonstrate that the previously proposed $T\ln (t/\tau _0)$ 
scaling of the relaxation data at zero field \cite
{Vincentjpe94,Labartaprb93,Iglesiasjm95,Iglesiaszpb96,Balcellsprb97,Barbarajm93,Wernsdorferjm95}
can also be extended to systems relaxing in an applied magnetic field. We will show how information about the 
microsopic energy barriers responsible for the relaxation can be obtained 
from this procedure.

The paper is organized as follows. In Sec. \ref{model}, we present the basic
features of model showing how the distribution of energy barriers of the system
is influenced by the application of a magnetic field with the help of the
concept of effective energy barrier distribution. In Sec. \ref{relaxation}, 
we derive the equation governing the time dependence of the magnetization
from a master rate equation for a Two-State Approximation (TSA). 
The results of numerical calculations based on the above mentioned equation 
are presented in Sec.\ref {numerical}. There, we present the $T \ln(t/\tau_0)$ scaling 
of relaxation curves at a given magnetic field, discussing its range of validity. We also
study the possibility of a scaling at different fields and fixed
temperature, and its applications. Finally in Sec. \ref{conclusions} we
resume the main conclusions of the article.

%_/_/_/_/_/_/_/_/_/_/_/_/_/_/_/_/_/_/_/_/_/_/_/_/_/_/_/_/_/_/_/_/_/_/_/_/_/_/_/ 

\section{Model}

\label{model}
%----------------------------FIG. 1------------------------------- 
\begin{figure}[htbp] 
\centering
\includegraphics[width= 0.8\textwidth]{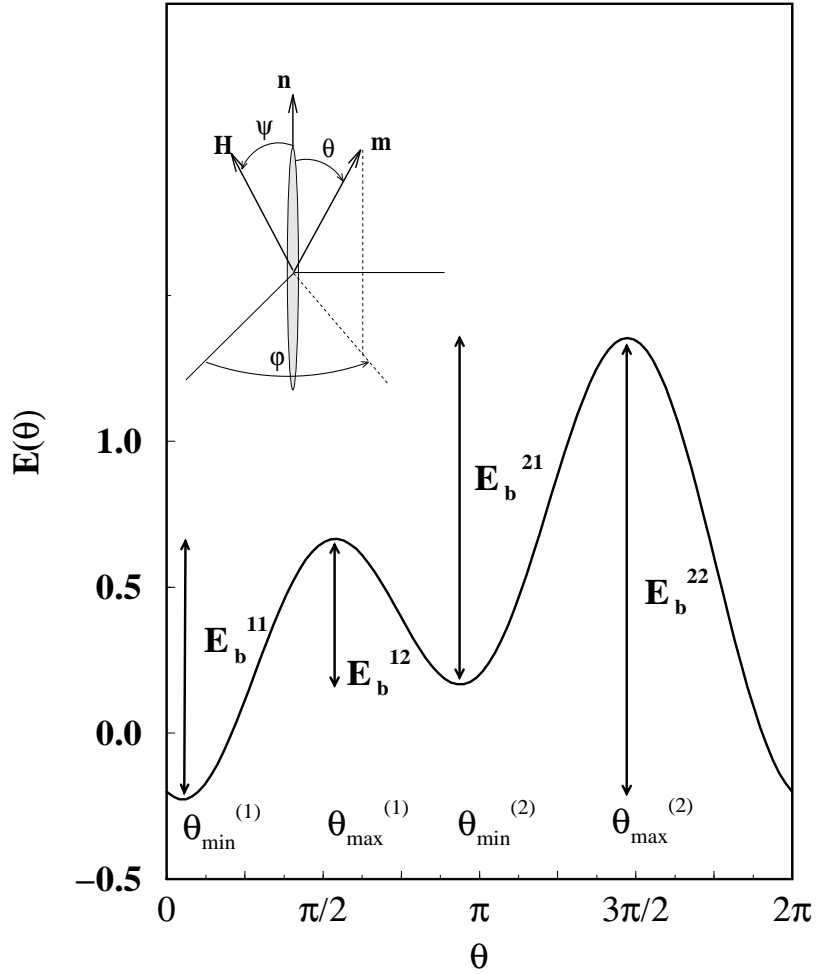} 
\caption{Energy function $E(\theta,\psi)$ as a function of the angle between 
the magnetization vector $\bf m$ and the magnetic field $\bf h$, for $\bf m$ 
in the plane formed by the easy-axis and the magnetic field ($\varphi=0$),
as given by Eq. (\ref{energy2}).  
The plot is for a particle whose easy-axis $\bf n$ forms an angle $\psi= 30^o$ 
with $\bf h$, and $H=0.3$. We have used the following  
notation to designate the extrema of the energy: 
$\theta^1_{\rm min}$ and $\theta^1_{\rm max}$ refer to the extrema closer to the 
field direction while $\theta^2_{\rm min}$ and $\theta^2_{\rm max}$ refer to  
those farther from the direction of the field. The four possible  
energy barriers between them are $E^{ij}_b\equiv E(\theta^i_{\rm max})- 
E(\theta^j_{\rm min})$.
Inset: Schematic representation of the quantities involved in the  
definition of the system. The easy-axis of the particles $\bf n$ are in
the x-z plane forming an angle $\psi$ with the magnetic field $\bf H$, 
which points along the z axis. $\theta$ and $\varphi$ are the spherical 
angles of the magnetization vector $\bf M$.  
} 
\label{energyplot} 
\end{figure} 
%----------------------------FIG. 1------------------------------- 
We consider an ensemble of randomly oriented, noninteracting single-domain
ferromagnetic particles of volume $V$ and magnetic moment ${\bf M}={M_s}V%
{\bf m}$ with uniaxial anisotropy. To take into account the spread of
particle volumes in real samples, we will assume that the particles
anisotropy constants $K$ are distributed according to a logatithmic-normal 
distribution
\begin{equation}  
\label{logn}
f(K)={\frac{1 }{{\sqrt{2\pi}K \sigma}}} e^{- \ln^2({K/K_0})/{2\sigma^2}} \ ,
\end{equation}
The energy of a particle is determined by the orientation of ${\bf M}$ with
respect to the external magnetic field ${\bf H}$ and to the easy-axis
direction ${\bf n}$. Using the angular coordinates defined in Fig. \ref
{energyplot}, it can be written as 
\begin{equation}  
\label{energy2}
\bar E=\frac{E}{V K}=-\cos^2(\theta)-2h\cos(\theta-\psi) \ .
\end{equation}
where we have defined the reduced field $h\equiv H/H_{{\rm c}%
}$ and $H_{{\rm c}}=2K/M_s$ as the critical field for an
aligned particle. We have concentrated on the two dimensional case (${\bf M}$
lying in the plane formed by ${\bf H}$ and ${\bf n}$) since the
energy maxima and minima can be calculated analytically only in this case.
In Fig. \ref{energyplot}, we show the variation of the energy with $\theta$
for a typical case, defining in the same figure the notation for the energy
barriers and extrema.

%_/_/_/_/_/_/_/_/_/_/_/_/_/_/_/_/_/_/_/_/_/_/_/_/_/_/_/_/_/_/_/_/_/_/_/_/_/_/_/ 
%----------------------------FIG. 2------------------------------- 
\begin{figure}[bp] 
\centering
\includegraphics[width= 1.0\textwidth]{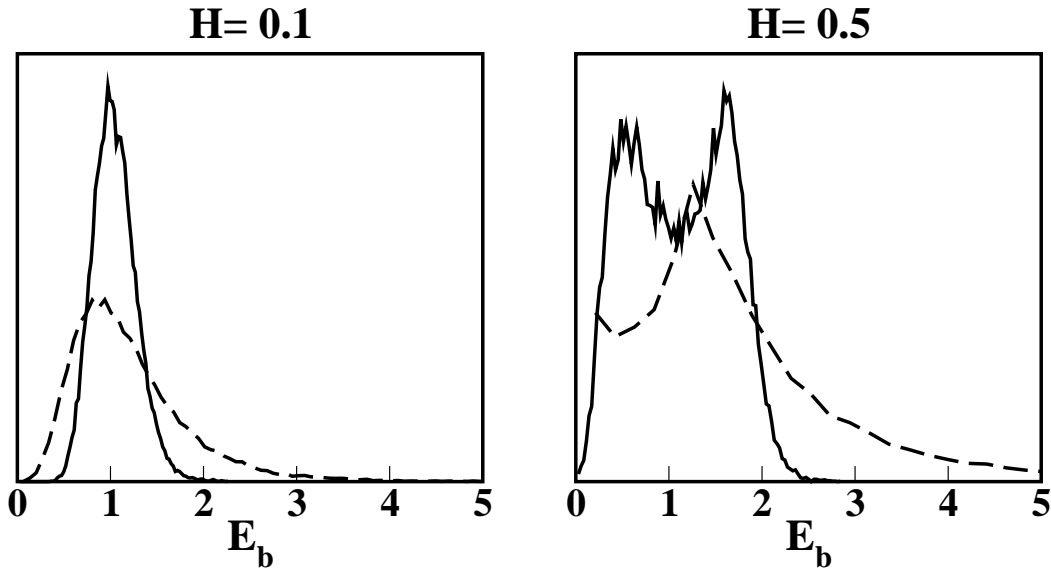} 
\caption{Effective energy barrier distributions for particles with
random orientations of anisotropy axes and a lognormal distribution 
of anisotropy constants of $\sigma= 0.2$ (continuous lines), $\sigma= 0.5$ (dashed lines), 
for values of $H$ as indicated in the figures. 
} 
\label{distriali} 
\end{figure} 
%----------------------------FIG. 2------------------------------- 
The magnetic field modifies the energy barriers of the system depending on
the particle orientation and anisotropy value, and, consequently, changes
the original energy barrier distribution \cite{Berkovjm92,Victoraprl91}. 
In order to understand the qualitative change of $f_{{\rm eff}}$ with $h$,
we have numerically calculated $f_{{\rm eff}}(E_{{\rm b}})$ for a system of
randomly oriented particles with $f(K)$,for different widths $\sigma$ and $K_0=1$, and several values of the magnetic field $h$. The calculation has been performed by making energy barriers histograms for a collection of 10~000 particles. The results are given in Fig. \ref{distriali}. In all the cases, we observe
the progressive splitting of the original distribution $f(E_{{\rm b}}^0)$ in
two subdistributions of high and low barriers as $h$ increases from zero.
The field tends to make deeper one of the minima, therefore increasing the
two energy barriers for rotation of ${\bf M}$ out of the field direction,
while the other two are reduced. In this way, the global effect of $h$ is a
splitting of $f(E_{{\rm b}})$ towards lower and higher values of $E_{{\rm b}}$.
As $h$ attains the critical value $h_{{\rm c}}$ for the particles with
smaller $K$, a peak of zero or almost zero energy barriers starts to appear
(see for example the curves for $H=0.5$ in the case $\sigma=0.5$);
while most of the non-zero barriers are distributed according to a
distribution identical to $f(E_{{\rm b}}^0)$, but centered at higher
energies. The higher the width of the distribution $\sigma$, the lower the $h
$ at which the lowest energy barriers start to be destroyed by the field.
The random orientation of the anisotropy axes smears out the distributions 
and makes the minima less pronounced than for a system with aligned particles
due to the spread in particle orientations, this effect being more noticeable at high $h$. 
In Sec. \ref{numerical}, we will discuss how these results affect the time
dependence of magnetization in relaxation experiments.
%_/_/_/_/_/_/_/_/_/_/_/_/_/_/_/_/_/_/_/_/_/_/_/_/_/_/_/_/_/_/_/_/_/_/_/_/_/_/_/ 
\section{Relaxation curves in the presence of a magnetic field}

\label{relaxation}

In order to derive an equation for the time dependence of the magnetization we 
propose a simple dynamical model that can be solved analytically in terms of 
intrinsic parameters. 
Since usually, in experiments performed in real samples the thermal energy 
$k_B T$ is smaller than the anisotropy energy $KV$, we will concentrate on the 
high energy barrier (low $T$) regime.
In this case, it is justified to consider only transitions between the two 
energy minima of each particle, and to truncate the continuum of possible
states for the magnetization to the two energy minima \cite{Pfeifferpsa90b,Pfeifferpsa90c}(TSA).
The master equation for $P_i$, the probability to find the 
magnetization of the particle in the minimum $i$, can be written as\cite{Reifbook} 
\begin{equation}
\frac{dP_{i}}{dt}=\sum_{k=1,2}\sum_{j\neq i}\left\{
w_{ji}^{(k)}P_{j}-w_{ij}^{(k)}P_{i}\right\} \ ,  \label{difeq}
\end{equation}
where $w_{ij}^{(k)}$ stands for the transition rate for a jump from the
state $i$ to the state $j$ separated by the maximum $k$ (see Fig. \ref
{energyplot}). Our choice for these quantities takes into account the energy barrier 
separating the initial minimum $i$ from the final state $j$ 
\begin{equation}
w_{ij}^{(k)}=\frac{1}{\tau _{ij}^{(k)}}=\frac{1}{\tau _{0}}e^{-E_{{\rm b}}^{ki}\beta },
\end{equation}
where $\tau _{0}^{-1}$ is the attempt frequency and $\beta=1/k_{B}T$.  

Taking into account the normalization condition $P_{1}+P_{2}=1$, one can
easily solve Eq. (\ref{difeq}) for $P_{1}$ and $P_{2}$ as a function of time 
\begin{eqnarray}
&&P_{1}(t)={\frac{1-e^{\beta \varepsilon }e^{-t/\tau }}{1+e^{\beta
\varepsilon }}}  \nonumber \\
&&P_{2}(t)={\frac{e^{\beta \varepsilon }(1+e^{-t/\tau })}{1+e^{\beta
\varepsilon }}}\ .
\label{p1p2}
\end{eqnarray}
The time-dependence of the system is thus characterized by an exponential
function with a single relaxation time $\tau $ that takes into account all
possible probability fluxes 
\begin{equation}
\tau ^{-1}=\tau _{0}^{-1}\left( e^{-\beta E_{{\rm b}}^{22}}+e^{-\beta E_{%
{\rm b}}^{12}}\right) \left( 1+e^{\beta \varepsilon }\right) \ ,
\end{equation}
where $\varepsilon =E_{{\rm min}}^{1}-E_{{\rm min}}^{2}$ is a measure of the
asymmetry of the energy function.

The time dependence of the magnetization of the particle is then finally
given by: 
\begin{eqnarray}
&&m(t;K,\psi )=\cos [\theta _{{\rm min}}^{1}(\psi )]P_{1}(t)+\cos [\theta _{%
{\rm min}}^{2}(\psi )]P_{2}(t)  \nonumber \\
&&  \nonumber \\
&=&\bar{m}_{TS}(K,\psi )+[m_{0}-\bar{m}_{TS}(K,\psi )]e^{-t/\tau (K,\psi )}.
\end{eqnarray}
In this equation, $\bar{m}_{TS}(K,\psi )$ is the equilibrium magnetization
in the TSA, and $m_{0}$ is the initial magnetization. If we
have an ensemble of randomly oriented particles and a distribution of
anisotropy constants $f(K)$, then the relaxation law of the magnetization is
given by 
\begin{equation}
m(t)=\int_{0}^{\infty }dKf(K)\int_{0}^{\pi }d\psi \ m(t;K,\psi )\ .
\label{mtime}
\end{equation}
This will be the starting point for all the subsequent numerical
calculations of the relaxation curves and magnetic viscosity. 
%_/_/_/_/_/_/_/_/_/_/_/_/_/_/_/_/_/_/_/_/_/_/_/_/_/_/_/_/_/_/_/_/_/_/_/_/_/_/_/ 
\section{Numerical calculations}
\label{numerical}
\subsection{Relaxation curves: $T \ln(t/\tau_0)$ scaling and normalization
factors}

In this section, we present the results of numerical calculations of the
magnetization decay based on Eq. (\ref{mtime}). 
For the sake of simplicity, we have assumed
zero initial magnetization $m_{0}=0$, so particles have initially their
magnetic moments at random and evolve towards the equilibrium state $m_{{\rm %
eq}}$. In the following, we will use dimensionless reduced variables for
temperature and time, defined as $T/T_0$ and $t/\tau_0$, with $T_0=E_0/k_{%
{\rm B}}$ and $E_0$ the value of the energy at which $f(K)$ is centered.
%----------------------------FIG. 3------------------------------- 
\begin{figure}[htbp] 
\centering
\includegraphics[width= 1.0\textwidth]{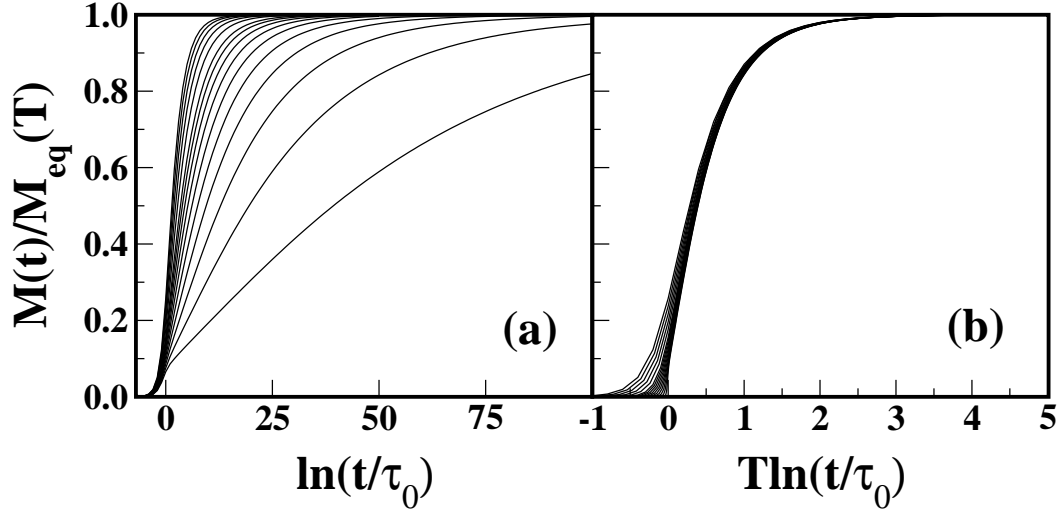} 
\caption{
(a) Relaxation curves for an ensemble of particles with randomly oriented 
anisotropy axes and the distribution $f(K)$ of Eq. \ref{logn}
of width $\sigma=0.5$ and $K_0=1$ calculated by numerical integration of 
Eq. (\ref{mtime}). 
Reduced temperatures $T/T_0$, starting from the lowermost curve, range from 
0.01 to 0.1 with 0.01 increments, and from 0.1 to 0.2 with 0.02 increments.
The applied field is $H= 0.5$. The original relaxation curves have been 
normalized to the equilibrium magnetization $m_{TS}(T)$.
(b) Relaxation curves of panel (a) plotted as a function of the scaling variable $\svar$
to obtain the master relaxation curve.
} 
\label{allrels05} 
\end{figure} 
%----------------------------FIG. 3------------------------------- 
In Fig. \ref{allrels05}, we show the results of the numerical calculations
for a system with $\sigma=0.5$ and $H= 0.5$ and temperatures ranging from 
$0.02$ to $0.2$. In the left hand panel, we
present the original relaxations normalized to the equilibrium magnetization
value that has be computed in the TSA. Normalization is essential in order to
compare relaxations at different temperatures \cite{Balcellsprb97},
especially at low fields where the temperature dependence of the equilibrium
magnetization is more pronounced.

Our next goal is to investigate the possibility of scaling relaxation curves
at different $T$ in a given magnetic field with the scaling variable $T
\ln(t/\tau_0)$, in the spirit of our previous works \cite
{Labartaprb93,Iglesiasjm95,Iglesiaszpb96,Balcellsprb97}. 
Let us briefly recall the basic points of the method. The time dependence of the remanent magnetization of a system with a distribution of energy barriers $f(E)$ can be written as
\begin{equation}
M(t)=M_0\int_0^{\infty} dE f(E)\ e^{-t/\tau(E)} \simeq M_0 \int_{E_{c}(t)}^{\infty} dE f(E)\ ,
\label{M(t)}
\end{equation}
with a characteristic relaxation time given by the Arrhenius-Ne\'{e}l theory as $\tau(E)=\tau_0\exp^{E/T}$.
Then, the relaxation law for the magnetization will be a function of the parameter $E_{c}(t)=\svar$ if the Boltzmann exponential factor in (\ref{M(t)}) can be approximated by a step function centered at $E_{c}(t)$. Then, as was shown in \cite{Labartaprb93}, the requirement for the validity of this approximation is that the characteristic width of $f(E)$ is of the order of the width of the step-like function $e^{-t/\tau(E)}$, which can be calculated as $\lambda=eT$ ($e$ is the Napier number). This implies low enough $T$ or energy distributions spread over a wide range of energies (large $\sigma$).

In Fig. \ref{allrels05}b, we show the original relaxation curves of Fig. \ref{allrels05}a as a function of the scaling variable $T \ln(t/\tau_0)$. The collapse of the curves into a single master curve indicates that scaling is accomplished also in the presence of a magnetic field. Only at short times, the higher $T$ curves depart from the master curve because in this $T$ range the scaling requirements are no longer accomplished.
According to the requirements for $\svar$ scaling presented in the preceding paragraph, in the presence of a magnetic field, the validity of the scaling hypotehsis at a given $T$ depends on the width of the effective energy barrier distribution (see Fig. \ref{distriali}). 
From simulations of master curves obtained at different $H$ (not presented in Fig. \ref{allrels05}), we observe that, the higher the field, the better the scaling of the curves is in the long time region and
the worse at short times. This observation holds independently of the value
of $\sigma$, indicating that it is a consequence of the application of a
magnetic field. This can be understood with the help of the effective energy
barrier distribution introduced in Sec. \ref{model}. As was shown in Fig. 
\ref{distriali}, $H$ widens $f_{{\rm eff}}(E)$ and shifts the lowest energy
barriers towards the origin, giving rise to a subdistribution of almost zero
energy barriers that narrows with increasing $H$, and, consequently, the requirements
for $T \ln(t/\tau_0)$ scaling are worse fulfilled at small $T \ln(t/\tau_0)$
values. On the contrary, as itwas shown in section \ref{model}, $H$
broadens the high energy tail of energy barriers that contribute to the
relaxation, $f(E_{{\rm b}}^{22})$, improving the scaling requirements at
large $T \ln(t/\tau_0)$ values.
%_/_/_/_/_/_/_/_/_/_/_/_/_/_/_/_/_/_/_/_/_/_/_/_/_/_/_/_/_/_/_/_/_/_/_/_/_/_/_/ 
\subsection{Scaling of relaxation curves at different magnetic fields}

Another interesting point is the possibility of finding an appropriate
scaling variable to scale relaxation curves at different fields for a given $%
T$, in a way similar to the case of a fixed field and different
temperatures, in which $T \ln(t/\tau_0)$ is the appropriate scaling
variable. In a first attempt, we will study the effect of $h$ on a system
with random anisotropy axes and distribution of anisotropies $f(K)$ of
width $\sigma$, Eq. (\ref{logn}). 
%----------------------------FIG. 4------------------------------- 
%\begin{figure}[htbp] 
\begin{figure}[bp] 
\centering
\includegraphics[width= 1.0\textwidth]{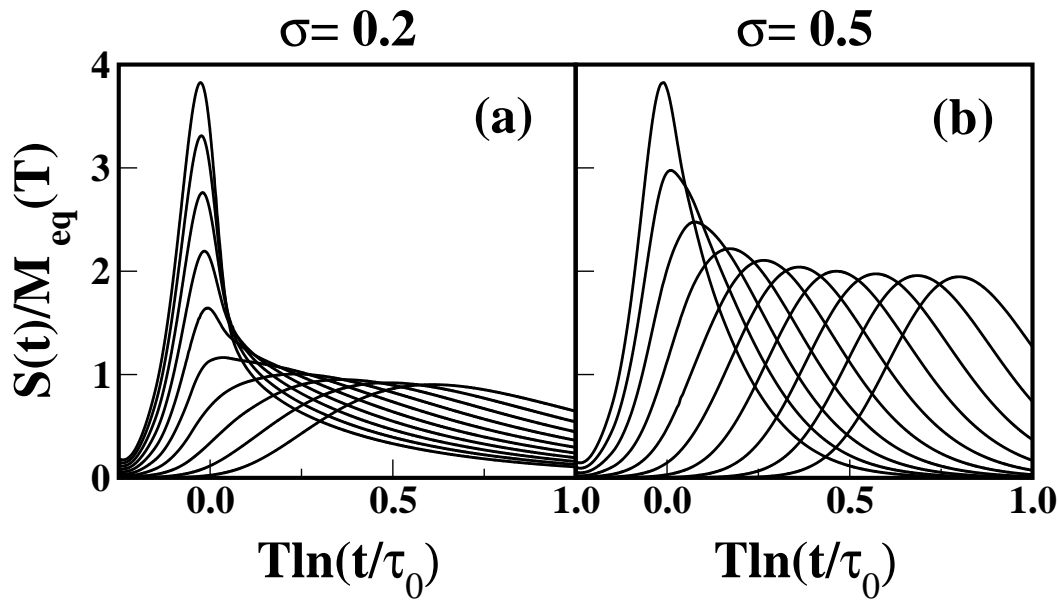} 
\caption{Low temperature ($T=0.05$) viscosity curves for a system of 
particles with random orientations and lognormal distribution 
of anisotropies with (a) $\sigma= 0.2$ and (b) $\sigma= 0.5$.
The curves have been normalized to the equilibrium magnetization and 
correspond to magnetic fields $H= 0.1$ to $1.0$ in 0.1 steps 
starting from the right.
} 
\label{sig02t_1.ali} 
\end{figure} 
%----------------------------FIG. 4------------------------------- 
Low temperature relaxation rates,
obtained by performing the logarithmic time derivative of $M(t)$
\beq 
\label{St} 
S(t)= \frac{dM(t)}{d[\ln(t)]}\simeq Tf(E_c(t,T)) \ , 
\eeq 
which is the coeficient of magnetic viscosity $S(t)$ are presented in 
Fig. \ref{sig02t_1.ali}, for $\sigma= 0.2, 0.5$. From this expression, we see that when $S$ is evaluated by taking the derivative of the master relaxation curve with respect to $\svar$, the resulting curve is proportional to the effective energy barrier distribution. 
%----------------------------FIG. 5------------------------------- 
\begin{figure}[htbp] 
\centering
\includegraphics[width= 1.0\textwidth]{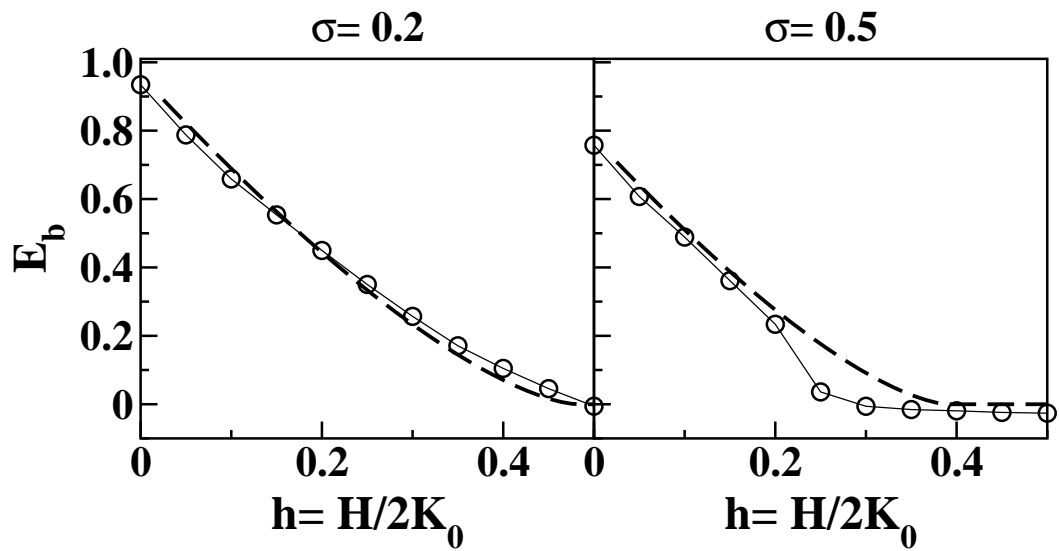} 
\caption{Field dependence of the energy corresponding to the maximum 
relaxation rate, $T\ln(t_{max}/\tau_0)$, as derived from the viscosity curves
in Fig. \ref{sig02t_1.ali} for $T=0.05$ (circles). The dashed line shows 
the $h$ dependence of the lowest energy barriers for the particles oriented 
at $\psi=\pi/4$ and $K$ equal to the most probable value.
} 
\label{smaxsig02.ali} 
\end{figure} 
%----------------------------FIG. 5------------------------------- 
First, let us notice that the qualitative shape of the viscosity curves is not distorted by $H$. 
The magnetic field simply shifts the position of the maxima towards lower
values of $T \ln(t/\tau_0)$ and narrows the width of the peaks, these
effects being similar for both studied $\sigma$.
The energy corresponding to the maximum relaxation rate ($E_{b}^{max}$) decreases with increasing 
$h$, following the decrease of the smallest energy barriers (see Fig. \ref
{smaxsig02.ali}), with an almost linear dependence on $h$. 
As can be clearly seen in Fig. \ref{smaxsig02.ali}, the dependence of $E_{{\rm max}}$ on $h$ (in symbols) follows that of the lowest energy barriers for particles oriented at $\psi= \pi/4$ and with 
$K=K_{{\rm max}}$ (dashed lines), which have been evaluated numerically from Eq. \ref{energy2}.
Let us also notice that $S_{{\rm max}}$ becomes almost constant below $h_0$ 
(the value at which the field starts to destroy the lowest energy barriers)
and low enough $T$ (see the position of the maxima in the curves of Fig. \ref {sig02t_1.ali}). 
%----------------------------FIG. 6------------------------------- 
\begin{figure}[htbp] 
\centering
\includegraphics[width= 1.0\textwidth]{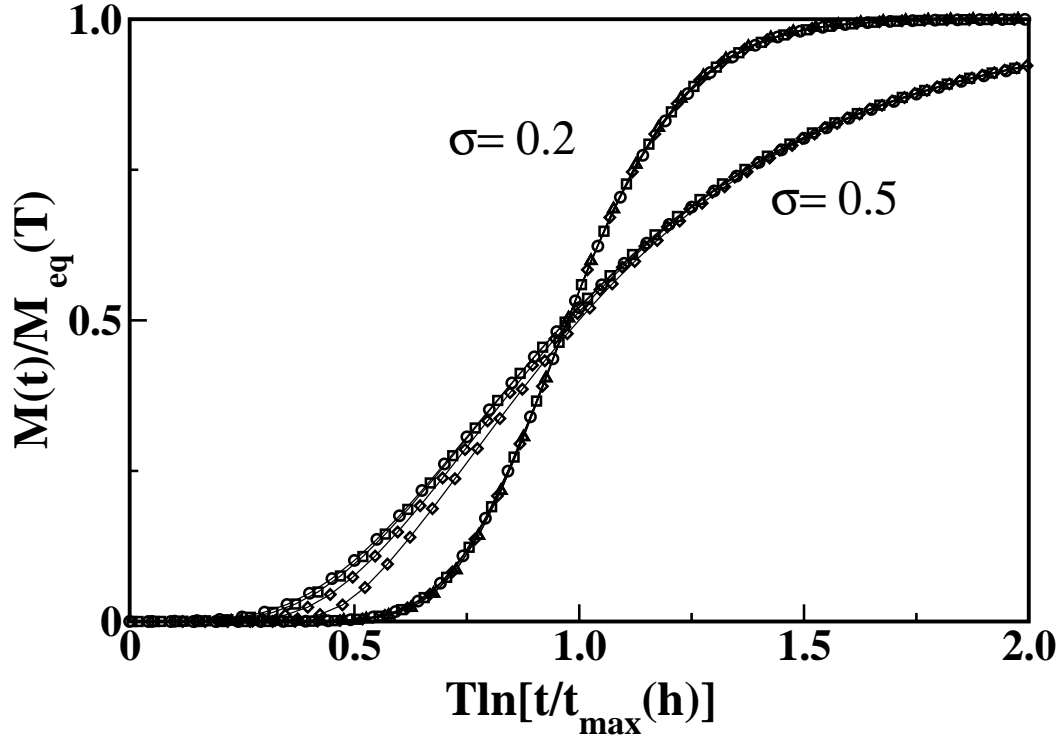} 
\caption{Normalized relaxation curves as a function of the scaling variable 
$T\ln[(t/t_{max}(h)]$ for $T= 0.05$) obtained from Fig. \ref{sig02t_1.ali}
by shifting the curves in the horizontal axis with the position of 
the maximum relaxation rate of Fig. \ref{smaxsig02.ali}.
For $\sigma= 0.2$ the field are $H= 0.1, 0.2, 0.3, 0.4, 0.5$, 
and for $\sigma= 0.5$, $H= 0.1, 0.2, 0.3, 0.4$ 
(starting from the uppermost curve).} 
\label{fig13} 
\end{figure}
%----------------------------FIG. 6------------------------------- The variation of $E_{b}^{max}$ and $S_{{\rm max}}$ with $h$, 
indicates that the relaxation curves at different $h$ and fixed $T$ may be brought 
to a single curve by shifting
them along the $T \ln(t/\tau_0)$ axis in accordance to the $E_{{\rm max}}$
variation. The resulting curves are displayed in Fig. \ref{fig13} for $%
\sigma = 0.2, 0.5$. They are the equivalent of the master curves of Fig. \ref
{allrels05} for a fixed $h$ and different $T$. Now the appropriate scaling
variable is 
\begin{equation}
E_{{\rm sca}}=T\ln[t/t_{{\rm max}}(h)] \ ,
\end{equation}
which generalizes the scaling at fixed $T$ (another approach to a similar scaling law may be found in \cite{Lyberatosjp97}). This new scaling is valid for
fields lower than $h_0$, the field at which the lowest barriers start to be
destroyed and above which the relaxation becomes dominated by almost zero
energy barriers. Thus, as already discussed in the previous paragraphs, the
wider $\sigma$, the smaller the $h$ range for the validity of field scaling.

%_/_/_/_/_/_/_/_/_/_/_/_/_/_/_/_/_/_/_/_/_/_/_/_/_/_/_/_/_/_/_/_/_/_/_/_/_/_/_/ 

\section{Conclusions}

\label{conclusions}

We have proposed a model for the relaxation of small particles systems under
a magnetic field which can be solved numerically from analytical expressions and which allows to study
the effect of the magnetic field on the energy barrier distribution. In
particular, we have shown that the original energy barrier distribution
$f(E_{{\rm b}})$ is split
into two subdistributions which evolve towards higher and lower energy
values, respectively, as $h$ increases. 

For fields smaller than the critical values for the smallest barriers, the
relaxation curves at different $h$ and fixed $T$ can be collapsed into a
single curve, in a similar way than $T \ln(t/\tau_0)$ scaling for curves at
fixed $h$. Whereas the latter allows to extract the barrier distribution by
differentiation of the master curve \cite{Balcellsprb97}, the shifts in the $%
T \ln(t/\tau_0)$ axis necessary to produce field scaling, give the field
dependence of the mean relaxing barriers, a microscopic information which
cannot easily be inferred from other methods \cite{Sappeyprb97}. 
%_/_/_/_/_/_/_/_/_/_/_/_/_/_/_/_/_/_/_/_/_/_/_/_/_/_/_/_/_/_/_/_/_/_/_/_/_/_/_/ 

\section*{Acknowledgements}

We acknowledge CESCA and CEPBA under coordination of C$^4$ for the computer
facilities. This work has been supported by SEEUID through project
MAT2000-0858 and CIRIT under project 2000SGR00025.

%_/_/_/_/_/_/_/_/_/_/_/_/_/_/_/_/_/_/_/_/_/_/_/_/_/_/_/_/_/_/_/_/_/_/_/_/_/_/_/ 

%_/_/_/_/_/_/_/_/_/_/_/_/_/_/_/_/_/_/_/_/_/_/_/_/_/_/_/_/_/_/_/_/_/_/_/_/_/_/_/ 

%\input{Figs_COSTP3}

\end{document}